\documentclass{cimento_arxiv}

\usepackage{graphicx}  

\title{Spectroscopy of hadrons with heavy quarks from lattice QCD}
\author{Sasa~Prelovsek }
\instlist{\inst{}Faculty of Mathematics and Physics, University of Ljubljana, Slovenia
  \inst{}   Jozef Stefan Institute, Ljubljana, Slovenia }

\begin{document}

\maketitle

\begin{abstract}
 Lattice QCD results on   hadrons with heavy quarks are briefly  reviewed.  The focus is on the spectrum of  conventional and exotic hadrons. Structure of certain conventional hadrons is addressed as well.  
\end{abstract}

\section{Introduction}
\label{intro}

 Experiments revealed that hadrons  with the following minimal contents exist:  mesons $\bar qq$, baryons $qqq$, tetraquarks $\bar qq\bar qq$, pentquarks $\bar qqqqq$ and hybrid mesons $\bar qGq$.  The first two sectors correspond to  conventional hadrons, while the last three are referred to as exotic hadrons.  We will briefly review  lattice results on some of these states \footnote{Only few references are cited due to the page limit. Other references are listed in the slides.  }.

\vspace{0.2cm}

  \section{\bf Hadron spectroscopy with Lattice QCD}
 
 The {\it spectrum of hadrons} (below, near, or above threshold) is extracted from the energies $E_n$ of QCD eigenstates $|n\rangle$ on a finite and discretized lattice in Eucledian space-time. The eigen-energies $E_n$ are determined from the time-dependence of  two-point correlation functions 
$\langle O_i (t_E) O_j^\dagger (0)\rangle$
 $=\sum_{n} \langle O_i|n\rangle  ~e^{-E_n t_E} \langle n|O^\dagger_j\rangle~,$
  where operators $O$ create/annihilate the hadron system with a given quantum number of interest.   
    
    The masses of  \underline{strongly stable hadrons well below threshold} are obtained as $m\!=\!E_n\vert_{\vec p = 0}$. These have  already been determined  and  agree well with the experiment, e.g.  \cite{Hatton:2021dvg,Koponen:2022hvd}. 
 
 \begin{figure}[htb]
\centering
\includegraphics[width=0.90\textwidth,clip]{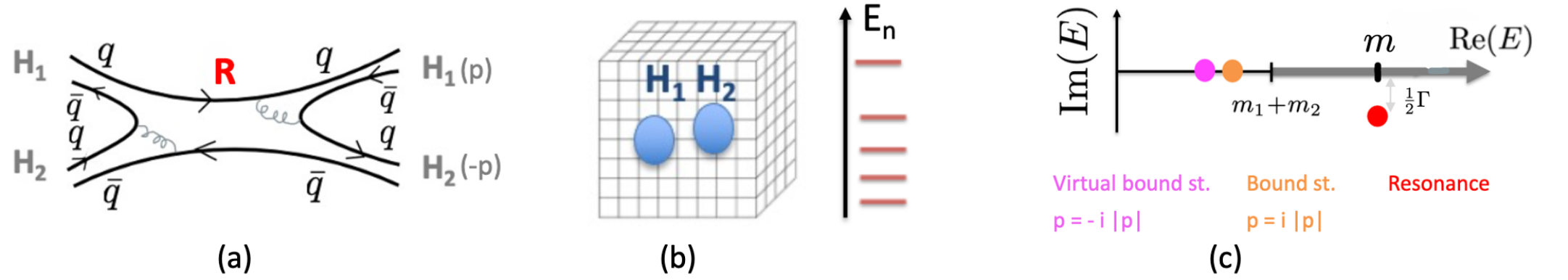}  
\caption{ Extracting   resonances and near-threshold bound states from one-channel scattering. }
\label{fig:1}
\end{figure}

 The masses of \underline{hadrons near threshold and hadronic resonances} have to be inferred from the scattering of two hadrons $H_1H_2$, which is encoded in the    scattering amplitude $T(E)$.    The simplest example is a one-channel  scattering  in partial wave $l$ sketched in Fig. \ref{fig:1}.     L\"uscher has shown that the energy  $E$ of a two-hadron eigenstate in finite volume    renders $T(E)$  at that energy in infinite volume  \cite{Luscher:1991cf}.  This relation  leads  to $T(E)$  for real $E$, which is then analytically continued to  complex energies. A pole  in $T(E)$ indicates  the presence of a state, while its position renders its mass $m=\mathrm{Re}(E)$ and the width $\Gamma=-2~\mathrm{Im}(E)$. A resonance corresponds to a pole away from the real axes. A bound state corresponds to a pole below the threshold: the state is referred to the bound state if the pole occurs for positive imaginary momenta $p=i|p|$ and a virtual bound state if it occurs for $p=-i|p|$, where $p$   denotes the magnitude of the 3-momentum in the center-of-mass frame. The majority of the scattering studies are still performed at $m_\pi>m_\pi^{phy}$ and at a single lattice spacing.

\begin{figure}[htb]
\centering
\includegraphics[width=0.80\textwidth,clip]{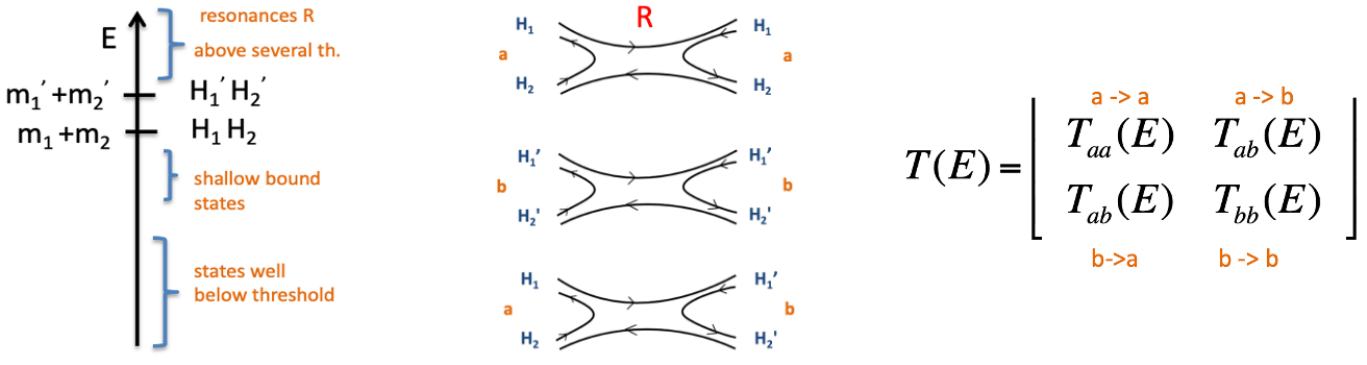}  
\caption{ A resonance that  decays via two  channels $R\to H_1H_2,~H_1^\prime H_2^\prime$   has to be inferred from the scattering of two coupled channels.    }
\label{fig:3}
\end{figure}
 
 The resonances that  decay via several strong decay channels $R\to H_1H_2,~H_1^\prime H_2^\prime, ..$ have to be extracted from the coupled-channel scattering (Fig. \ref{fig:3}).  The scattering matrix for two coupled channels contains three unknown functions of energy.  It is customary to parametrize their energy dependence in order to extract them from the eigen-energies using the L\"uscher's formalism. 
 
 The formalism and progress in addressing decays $R\to H_1H_2H_2$ was reviewed  at the last edition of this conference \cite{Romero-Lopez:2022usb}.

\section{Spectroscopy of various hadron sectors}

The majority of  the discovered exotic hadrons contain heavy quarks as those are more likely to form quasi-bound states due to small kinetic energies. Most of them decay strongly and the theoretical challenge to study them increases with the number of decay channels.

\begin{figure}[htb]
\centering
\includegraphics[width=0.9\textwidth,clip]{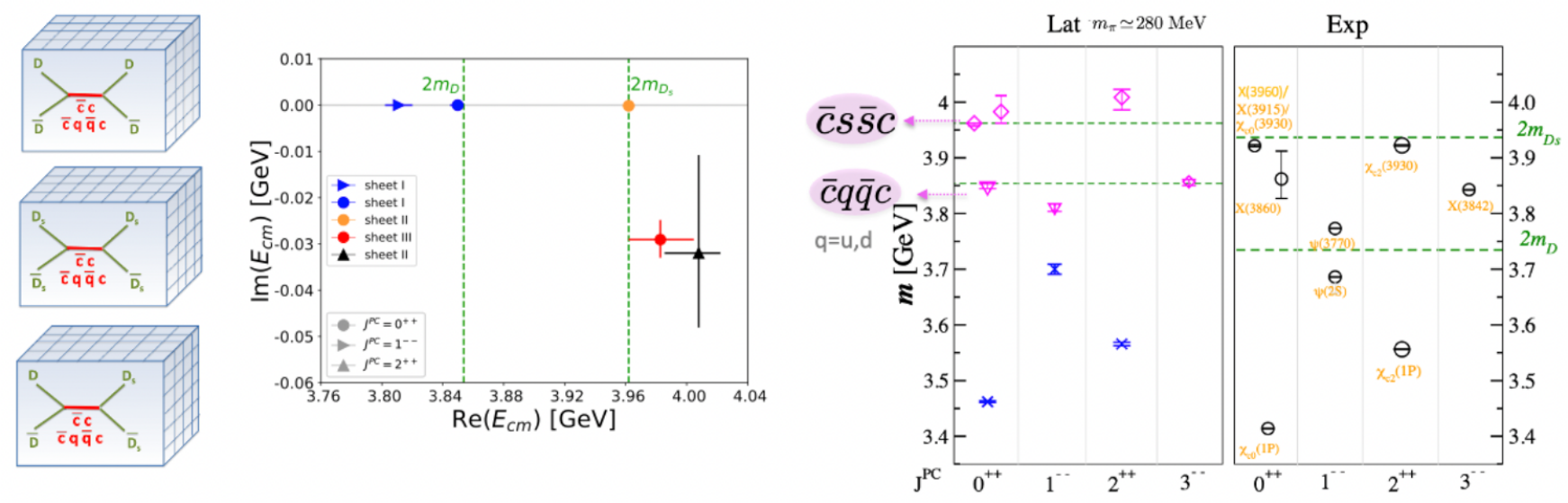}  
\includegraphics[width=0.8\textwidth,clip]{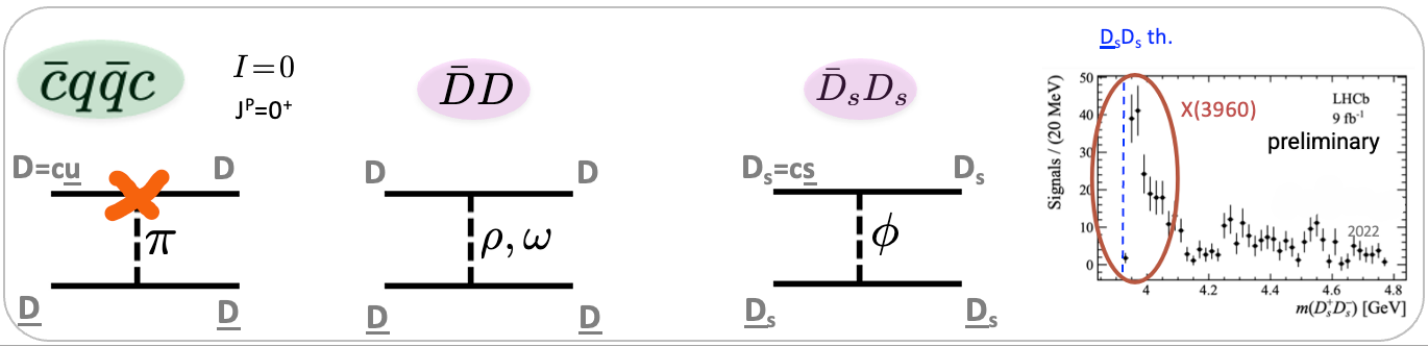} 
\caption{   Poles and masses for a charmonium-like system with $I=0$ from lattice   \cite{Prelovsek:2020eiw}. LHCb discovery of a $X(3960)$ \cite{LHCb:2022vsv} composed of $\bar cs\bar sc$. Possible binding mechanism.  }
\label{fig:4}
\end{figure}

\vspace{0.2cm}

{\bf  Charmonium-like states $\mathbf{\bar cc}$, $\mathbf{\bar cc\bar qq^\prime}$, $\mathbf{\bar ccuud}$}

\vspace{0.1cm}

The spectrum of charmonium-like  states with $I=0$ was extracted from the coupled channels $\bar DD-\bar D_sD_s$  at $m_\pi\simeq 280~$MeV \cite{Prelovsek:2020eiw} (Fig. \ref{fig:4}). All the states except for two (indicated by magenta arrows) appear to be conventional charmonia $\bar cc$ . In addition, two exotic scalar states are predicted near both thresholds. The heavier  one has a large coupling to  $\bar D_s D_s$ and a small coupling to $\bar DD$ - it likely corresponds to a state $X(3960)$ composed of $\bar cs\bar sc$  recently discovered by LHCb \cite{LHCb:2022vsv}.  The two additional scalars were not found at $m_\pi\simeq 390~$MeV in \cite{Wilson:2023anv}. 

\begin{figure}[htb]
\centering
\includegraphics[width=0.99\textwidth,clip]{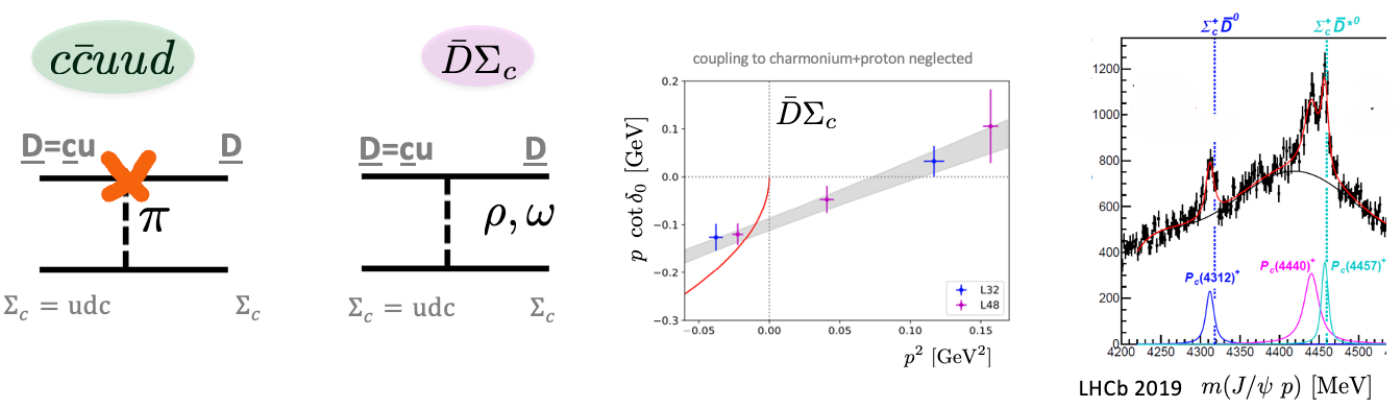}  
\caption{   $\bar D \Sigma_c$  scattering in $P_c$ channel  \cite{Xing:2022ijm}, LHCb discovery and possible binding mechanism.    }
\label{fig:Pc}
\end{figure}

A  candidate for pentaquark $P_c=\bar ccud$ with $J^P=1/2^-$ was found $6\pm 3~$MeV below threshold  $\bar D \Sigma_c$ threshold \cite{Xing:2022ijm}. It appears as a bound state in  one-channel scattering amplitude  $\bar D \Sigma_c$ (Fig. \ref{fig:Pc}), where the lower-lying channel $J/\psi p$ was omitted. 
 
\vspace{0.2cm}

{\bf Hadrons with two heavy quarks:  $\mathbf{QQ\bar q\bar q^\prime}$  }

\vspace{0.1cm}

\begin{figure}[htb]
\centering
\includegraphics[width=0.3\textwidth,clip]{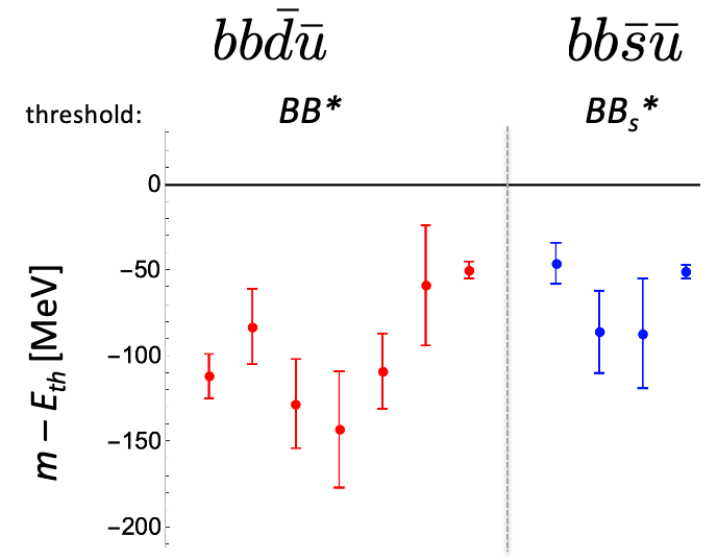}  $\quad $
\includegraphics[width=0.6\textwidth,clip]{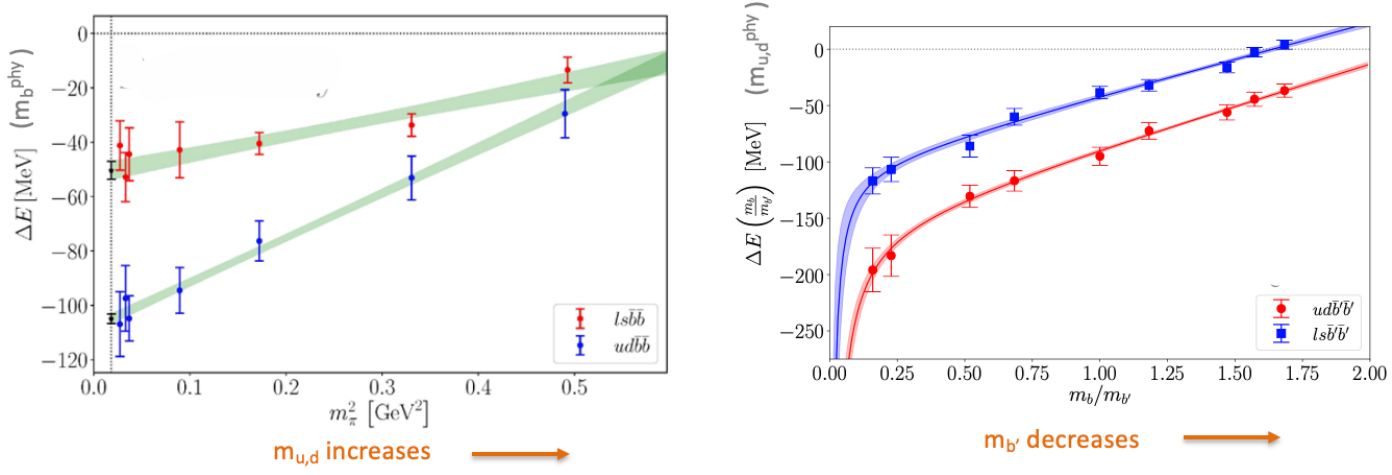}  
\caption{ Left: The binding energies of  $bb\bar u\bar d$ and $bb\bar u\bar s$ with  $J^P=1^+$ from  lattice (see references in the slides).     Right:  The dependence of the binding energy on $m_b$ and $m_{u,d}$ \cite{Francis:2018jyb,Colquhoun:2022dte}.  }
\label{fig:10}
\end{figure}

 The   $bb\bar u\bar d$ and $bb\bar s\bar d$ tetraquarks with $J^P\!=\!1^+$ are expected to reside  significantly below strong decay thresholds (Fig. \ref{fig:10}). This is a reliable conclusion based on a number of lattice simulations and model-based calculations (listed in the slides). The hadron $bb\bar u\bar d$  with such a deep binding  is expected to have a small size, which   indicates the dominance of  $[bb]_{\bar 3_c}^{S=1}[\bar u\bar d]_{3_c}^{S=0}$.   So far, this  the only tetraquark where lattice finds a strong support for the dominance of the diquark antidiquark Fock component.

\begin{figure}[htb]
\centering
\includegraphics[width=0.99\textwidth,clip]{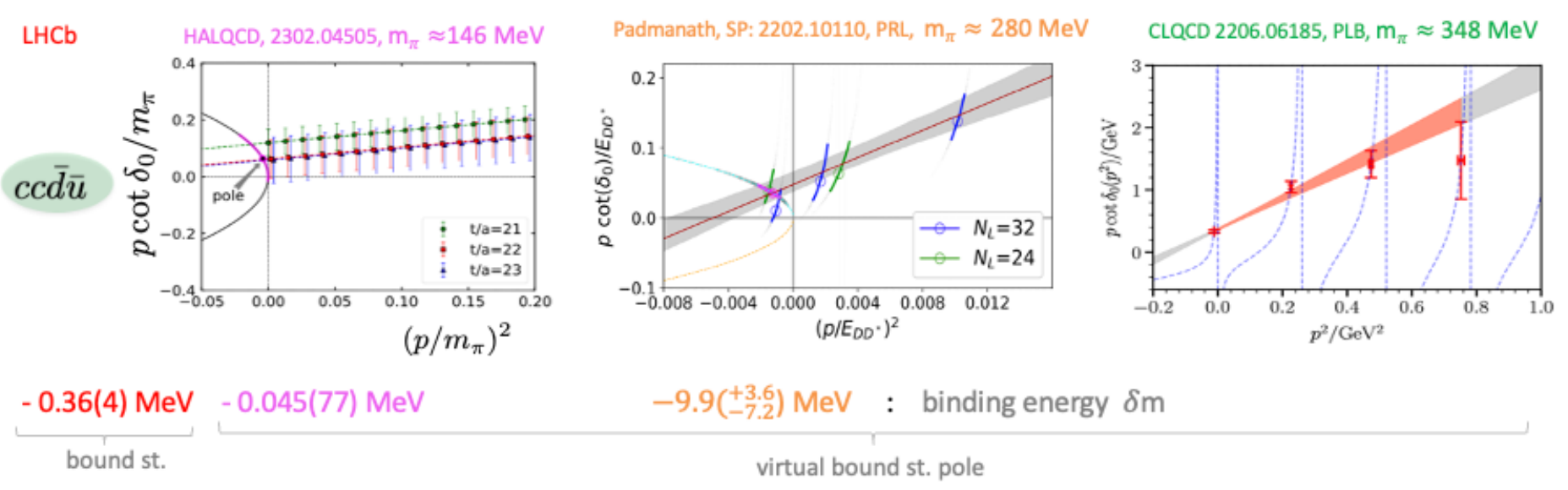}  
\caption{ The $DD^*$ scattering  in $T_{cc}$ channel and  the pole locations from lattice \cite{Lyu:2023xro,Padmanath:2022cvl,Chen:2022vpo}.  }
\label{fig:11}
\end{figure}

 Tetraquarks $QQ \bar q \bar q^\prime$ ($Q\!=\!c,b~, q\!=\!u,d,s$) with other flavors and spin-parities are expected near or above strong decay thresholds $H_1H_2$, as suggested also by Fig. \ref{fig:10}. In order to prove the existence of such a state, one needs to extract the scattering matrix and establish a pole in it  as shown in Fig. \ref{fig:1}.

The doubly charm tetraquark $T_{cc}=cc\bar u\bar d$ was discovered by LHCb  just $0.4$ MeV below $DD^*$ threshold  \cite{LHCb:2021vvq}, it has   $I\!=\!0$ and most likely $J^P\!=\!1^+$. The lattice results on $DD^*$ scattering from three simulations  at different $m_\pi$ are shown in Fig. \ref{fig:11}. All simulations find significant attraction; the attraction decreases with increasing $m_\pi$. This implies that $T_{cc}$ would-be  bound state from experiment converts to virtual bound state or a resonance at heavier $m_\pi$. Virtual bound state poles are indeed found in simulations \cite{Lyu:2023xro,Padmanath:2022cvl,Chen:2022vpo} when assuming effective range approximation. Relaxing this approximation and taking into account the effect of left-hand cut from one-pion exchange leads to a pair of virtual bound states or a resonance \cite{Du:2023hlu}. The long-range potential   is dominated by $\pi\pi$ exchange  in \cite{Lyu:2023xro}. 
The dominant attraction in this channel is attributed to $\rho$ exchange  in \cite{Chen:2022vpo}. 

\vspace{0.2cm}

{\bf Hadrons with a single heavy quark  }

\vspace{0.1cm}

\begin{figure}[htb]
\centering
\includegraphics[width=0.85\textwidth,clip]{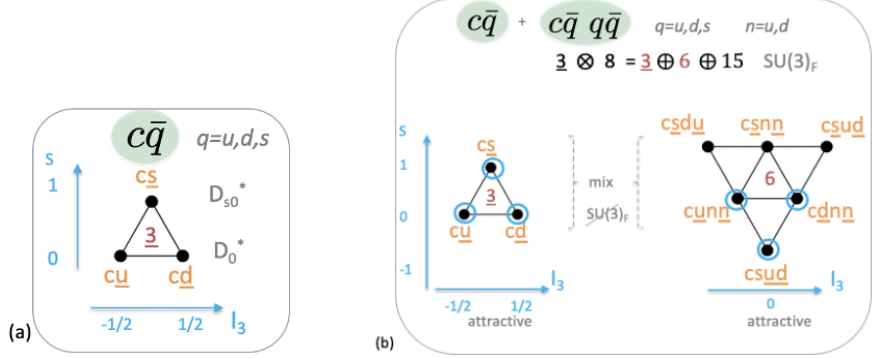}  
\caption{ Scalar charmed mesons according to the quark model (a) and the new paradigm (b).  }
\label{fig:12}
\end{figure}

The charmed scalar mesons  would form a $SU(3)$ flavor triplet in Fig. \ref{fig:12}a according to the quark model. However, a new paradigm is supported by the effective field theories based on HQET and ChPT, combined with the lattice results  as well as  the experimental data (see slides for many references).   According to this paradigm, the spectrum features $c\bar q$ as well as $c\bar q~\bar qq$ Fock components ($q=u,d,s$). The latter decomposes to the multiplets $\bar 3\oplus 6\oplus 5$ in the $SU(3)$ flavor limit.   The attractive interactions within the anti-triplet and the sextet suggest the existence of hadrons with flavors indicated by cyan circles in Fig. \ref{fig:12}b, with   two pair of poles for $I=1/2$ charmed mesons. The lower one resides at $2.1\!-\!2.2~$GeV in agreement with the lattice simulations and it is a natural partner of  $D_{s0}^*(2317)$.  The heavier pole at $2.4\!-\!2.5~$GeV is suggested by the EFT re-analysis of the lattice   and  experimental data.    

\begin{figure}[htb]
\centering
\includegraphics[width=0.5\textwidth,clip]{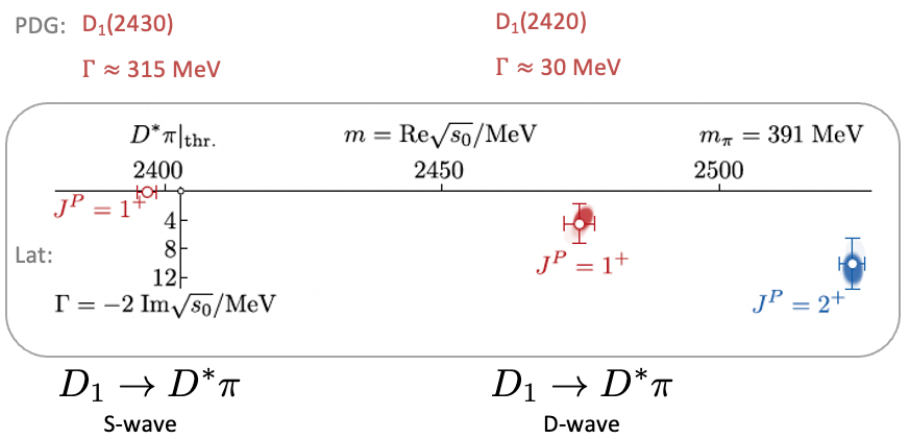}  
\caption{  Poles  related to charmed mesons with $J^P=1^+,2^+$ from \cite{Lang:2022elg}.  }
\label{fig:D1}
\end{figure}

The simulation of $D^*\pi$ scattering  finds one axial charmed meson dominated by s-wave and the other by  d-wave coupling to $D^*\pi$ \cite{Lang:2022elg} (Fig. \ref{fig:D1}). This is in line with HQET prediction and significantly  different experimental decay widths.

\vspace{0.2cm}

{\bf  Bottomonium-like states $\mathbf{\bar bb}$, $\mathbf{\bar bGb}$, $\mathbf{\bar bb\bar qq^\prime}$}

\vspace{0.1cm}

\begin{figure}[htb]
\centering
\includegraphics[width=0.99\textwidth,clip]{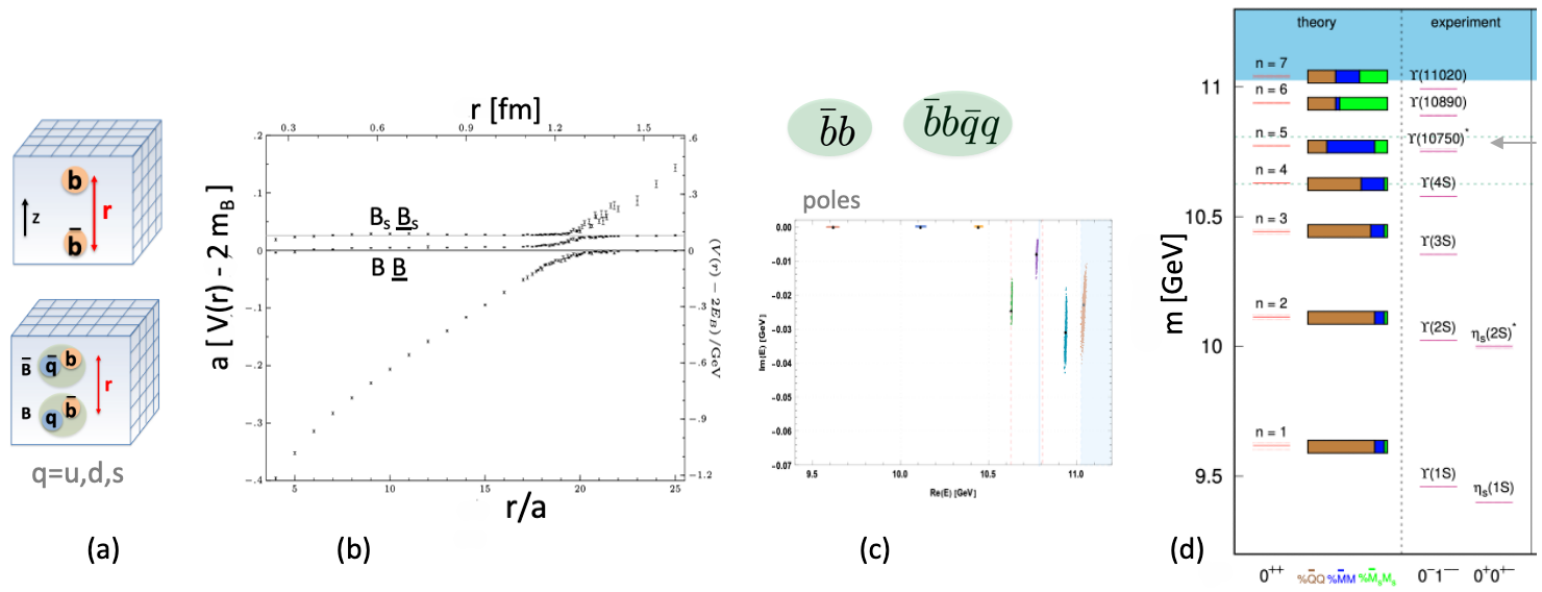}  
\caption{   (a,b) The static potential  of the quarkonium system that accounts also for the coupling to a pair of heavy mesons  \cite{Bulava:2019iut}. (c,d) Poles and composition of the bottomonium-like states \cite{Bicudo:2022ihz} from analogous static potential \cite{Bali:2005fu}.   }
\label{fig:6}
\end{figure}

\begin{figure}[htb]
\centering
\includegraphics[width=0.8\textwidth,clip]{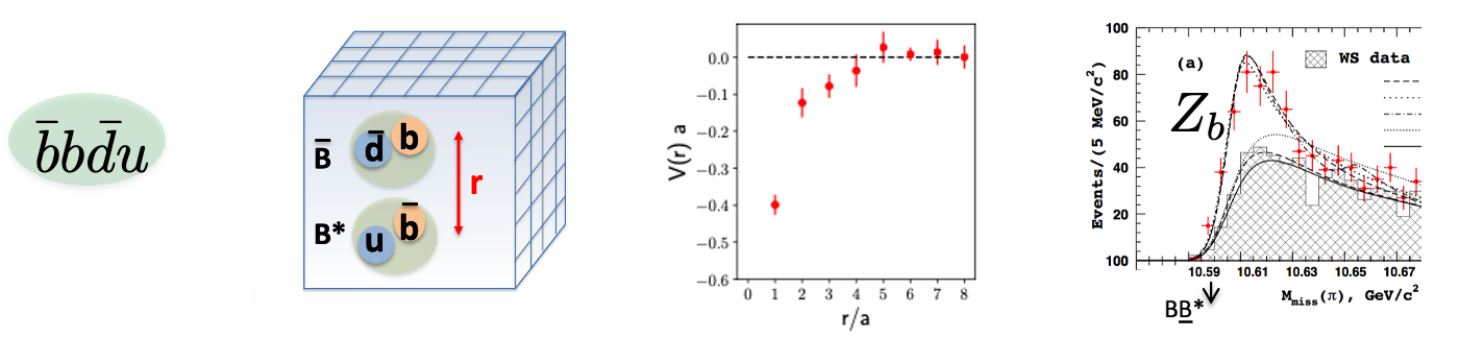}  
\caption{ The attractive static potential between $B$ and $\bar B^*$   \cite{Prelovsek:2019ywc}  is likely responsible for the existence of $Z_b\simeq  \bar bb\bar du$ \cite{Belle:2015upu}.     }
\label{fig:7}
\end{figure}

\begin{figure}[htb]
\centering
\includegraphics[width=0.8\textwidth,clip]{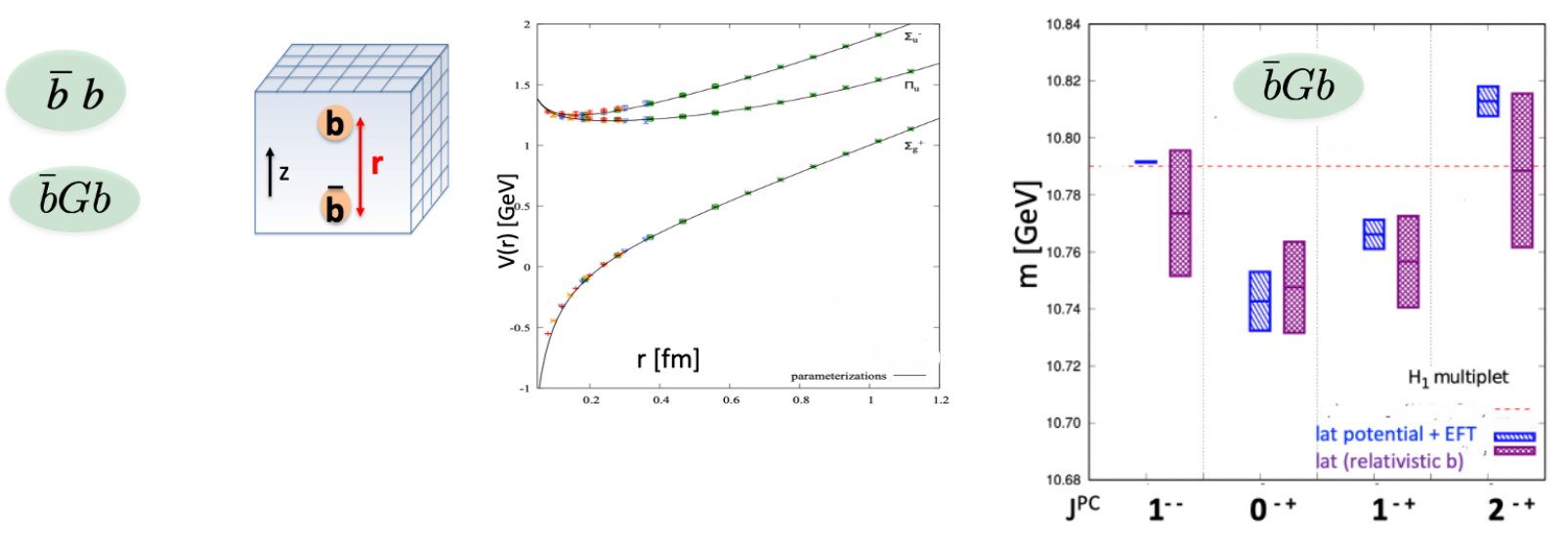}  
\caption{    The   potentials related to  hybrids  from \cite{Schlosser:2021wnr}. Masses of $\bar bGb$ hybrids (update of \cite{Brambilla:2018pyn,Brambilla:2019jfi}) from older potentials  \cite{Juge:1997nc}   (blue) and from relativistic $b$ quarks \cite{Ryan:2020iog} (violet).    }
\label{fig:8}
\end{figure}

 These systems can be studied using relativistic, non-relativistic  or static b-quarks. Here  I focus on the last option, where  two heavy quarks  and additional light degrees of freedom are investigated via the Born-Oppenheimer approximation. The eigen-energies at fixed distance between heavy quarks render the potential $V(r)$. Motion of heavy quarks within this potential is then studied with   Schr\"odinger-type equation. The aim is to  determine whether bound states or resonances exist.
 
   The static potential for $\bar bb$ system with $I=0$ that accounts also for the coupling to a pair of heavy mesons \cite{Bulava:2019iut}  is shown in Fig. \ref{fig:6}b.  Coupled-channel Schr\"odinger equation with analogous  potential (from an earlier calculation \cite{Bali:2005fu}) renders the poles related to bottomonium-like states  and their composition   in terms of $\bar bb$, $\bar BB$ and $\bar B_sB_s$  \cite{Bicudo:2022ihz} (Figs. \ref{fig:6}c,d).  These are dominated by the conventional Fock component $\bar bb$, except for state $n\!=\!5$,  which is likely related to unconventional $\Upsilon(10750)$ discovered by Belle. 
   
    The observed $Z_b$ resonances with flavor content $\bar bb\bar du$ are challenging for rigorous treatment since the lowest decay channel is $\Upsilon_b \pi$, while they reside at the higher threshold $B\bar B^*$. This was taken into account in  the extraction of the potential between $B$ and $\bar B^*$ in Fig. \ref{fig:7}, which is  attractive at small distances  \cite{Peters:2016wjm,Prelovsek:2019ywc,Sadl:2021bme}. This attraction is likely responsible for the existence of the exotic $Z_b$. 
   
    The excited potentials for $\bar bb$  with certain spin-parities in Fig. \ref{fig:8} are relevant for hybrid  mesons $\bar b Gb$. The   masses from these potentials within the Born-Oppenheimer approach  \cite{Brambilla:2018pyn,Brambilla:2019jfi} agree with those obtained using relativistic $b$-quarks  \cite{Ryan:2020iog}.

     \begin{figure}[htb]
\centering
\includegraphics[width=0.7\textwidth,clip]{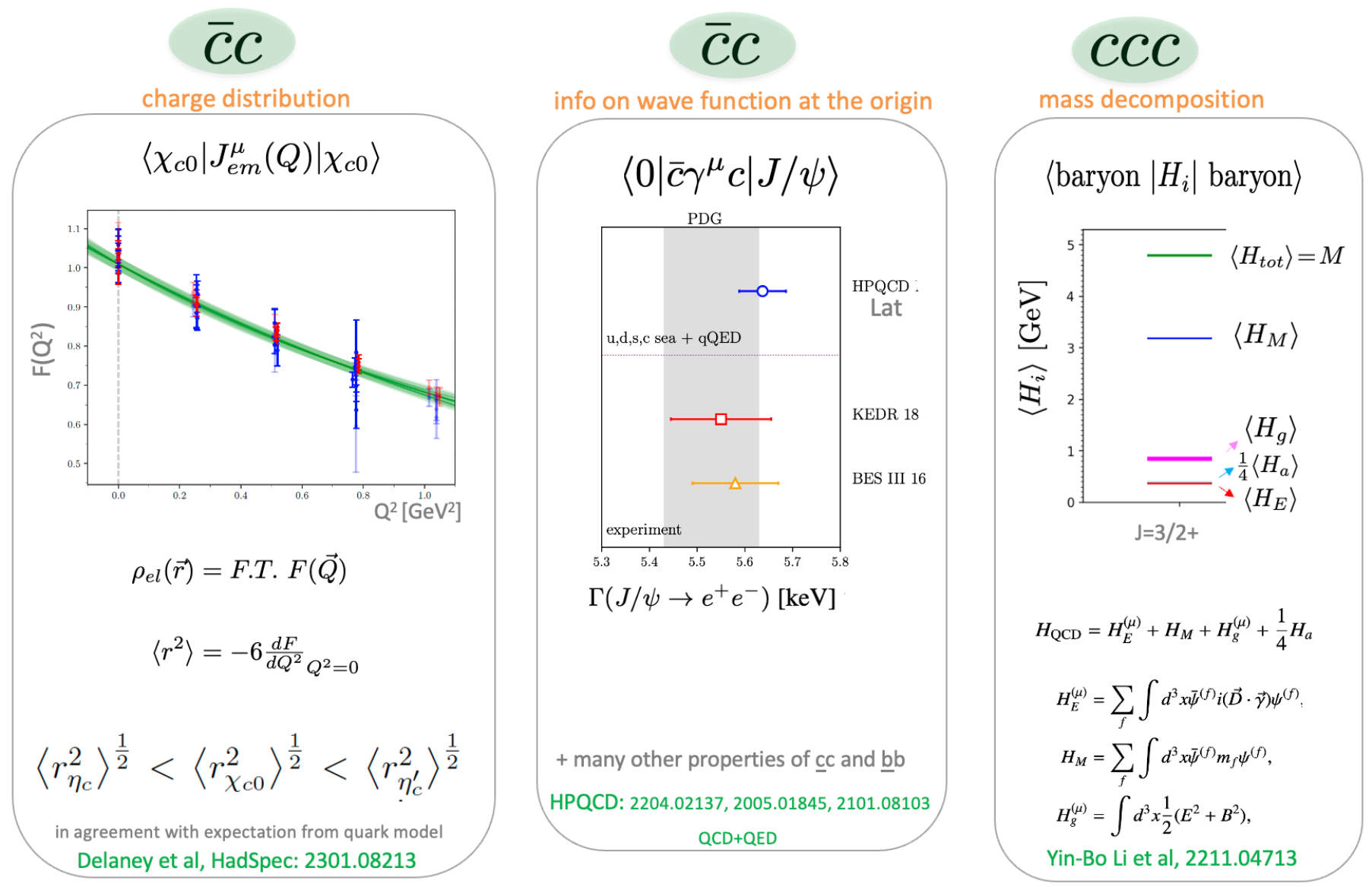}  
\caption{Probing structure of hadrons with charm quarks in simulations \cite{Delaney:2023fsc,Koponen:2022hvd,Li:2022vbc}.}
\label{fig:structure}
\end{figure}

 \section{Structure of hadrons with heavy quarks}

The charge distribution of charmonia was probed via EM and other currents in \cite{Delaney:2023fsc}, while various decay constants that probe the wave function were extracted in \cite{Koponen:2022hvd}. The mass decomposition with respect to various terms of the QCD Hamiltonian was determined for  charmed baryons  in \cite{Li:2022vbc}  (Fig. \ref{fig:structure}).

\section{Looking ahead} 

\begin{figure}[htb]
\centering
\includegraphics[width=0.8\textwidth,clip]{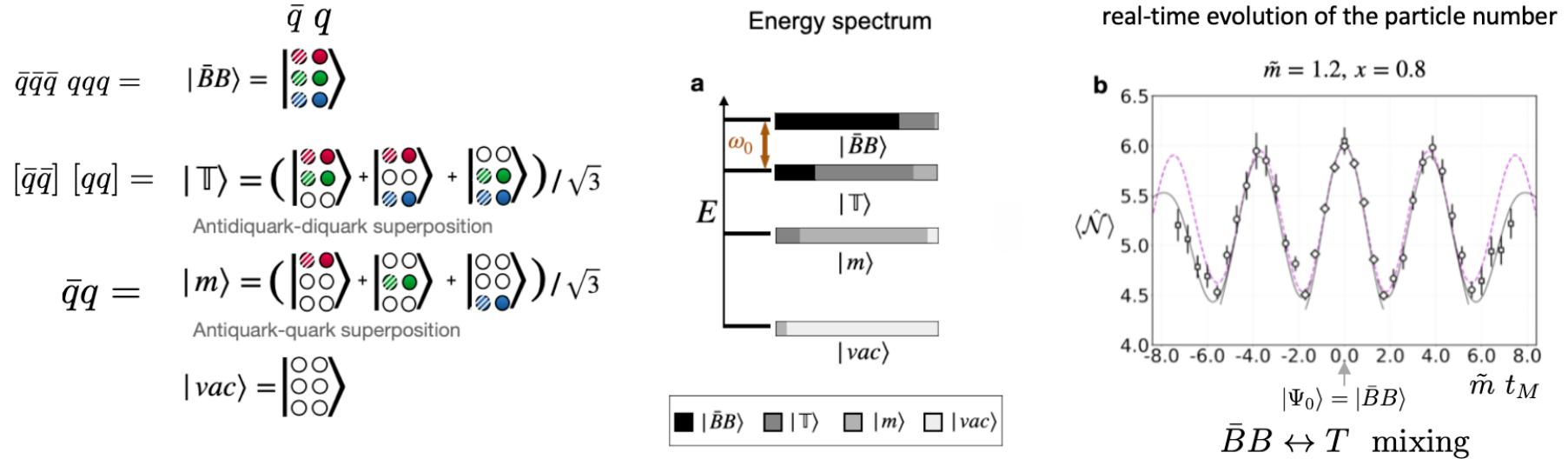}  
\caption{ Tetra- and penta-quarks in one-dimensional QCD from quantum computer \cite{Atas:2022dqm}.}
\label{fig:quantum-computer}
\end{figure}

The drawback of spectroscopy  on  Eucledian lattice is suppressed contribution $e^{-E_n t_{E}} $ of the excited states. The evolution of the QCD systems in real Minkovsky time  $e^{-iE_n t_{M}} $ is one of the long-term goals for quantum computers.  Such time evolution for tetraquark and pentquark systems has already been studied in one-dimensional QCD with one quark flavor on a quantum computer \cite{Atas:2022dqm} (Fig. \ref{fig:quantum-computer}). 

\section{Conclusions}

Experiments have provided great discoveries of new conventional as well as around thirty exotic hadrons. I have reviewed the theoretical challenge to understand the spectroscopic properties of 
various hadron sectors from lattice QCD. This  approach renders masses of  hadrons that are strongly stable, as well as most of the hadrons that are slightly below the strong decay threshold or decay strongly via one decay channel. The theoretical challenge increases with the number of open decay channels. It seems impossible to address  the high-lying states like $Z_c(4430)$ with current lattice methods, while many interesting physics conclusions are already available for certain lower-lying states. 

\vspace{0.4cm}

{\bf Acknowledgments}

 I acknowledge the  support from  ARRS research core funding No. P1-0035.


%

\end{document}